\documentclass[fleqn,10pt]{wlscirep}
\usepackage[utf8]{inputenc}
\usepackage[T1]{fontenc}
\usepackage{placeins}
\title{Sensorimotor features of a reversal learning task bias decision behavior without disrupting individual difference structure}

\author[1,2]{Elliot Huang}
\author[1]{William Xu}
\author[1,3,*]{Robert C. Wilson}
\affil[1]{School of Psychological and Brain Sciences, Georgia Institute of Technology, Atlanta, GA, USA}
\affil[2]{School of Computer Science, Georgia Institute of Technology, Atlanta, GA, USA}
\affil[3]{Center of Excellence for Computational Cognition, Georgia Institute of Technology, Atlanta, GA, USA}

\affil[*]{rwilson337@gatech.edu}

\begin{abstract}
In computational psychiatry, task-irrelevant factors such as a task's perceptual and motor features are typically assumed not to bias decision behavior, but at most to add noise. Yet growing evidence links sensorimotor processing to decision-making through multiple pathways, challenging this assumption. We tested this directly using a two-choice probabilistic reversal learning task completed by 90 participants under two sensorimotor conditions: a stationary condition requiring only arm movements to respond, and an active condition requiring participants to walk between physically separated computers. Task performance, measured as the propensity to choose the more probable rewarding option, did not differ between conditions. However, stay rate (the tendency to repeat the previous choice) was significantly elevated in the stationary condition, but only among participants who completed the active condition first; those who completed the stationary condition first showed no such difference. Logistic regression weights capturing the influence of recent win history on choice showed the same pattern, with stationary-condition elevation restricted to the active-first group. Cross-condition correlations for both measures were strong, indicating that this order-dependent effect shifted absolute measurements without disrupting individual difference structure. Prior sensorimotor history therefore does not compromise the task's ability to extract stable cognitive variables, but does bias the values it yields — a non-random source of variance that, if it generalizes beyond this experiment, could shift patients across diagnostic thresholds in proposed clinical applications. The finding adds to a growing list of potential methodological contaminants in computational psychiatry's cognitive-task paradigms that must be characterized and mitigated before clinical translation.
\end{abstract}
\begin{document}

\flushbottom
\maketitle
%
%
\thispagestyle{empty}


\section*{Introduction}

Computational psychiatry seeks to characterize the mind through the computations it uses to solve problems, with psychopathology distinguished as extreme values of the latent variables that parameterize these functions \cite{montague_computational_2012, adams_computational_2016, patzelt_computational_2018}. A primary method for estimating these latent variables is to design a task that requires the target computation, measure behavior repeatedly, and transform observations into proxies of the variables of interest. Such transformations range from simple behavioral metrics like accuracy or reaction time to fully specified computational models of the underlying process fit to trial-level stimuli and responses \cite{daw_trial-by-trial_2011, wiecki_model-based_2015, schwartenbeck_computational_2016, wilson_ten_2019}.

Human behavior is a function of many factors beyond the target latent variables and the designed task inputs (i.e., the provided feedback), including time of day \cite{mehrhof_both_2024}, transient mood states \cite{forgas_chapter_2017}, and task comprehension \cite{zorowitz_improving_2023}. What is necessary, then, for the cognitive-task approach in computational psychiatry to remain valid, is that these irrelevant factors must contribute only orthogonal residual variance: their effects cannot point in a particular direction, cannot correlate with the variable being measured, and cannot differ systematically between people or groups. Under these conditions they only add random noise, which can be counteracted simply by increasing sample size. Under standard consistency conditions, more observations allow estimation methods to converge on the true underlying values. If, however, an unknown contributor to behavior is structured, the same methods become problematic. What they converge to is no longer an estimate of the latent variable in isolation, but some mixture of the latent variable and the confound.

Sensorimotor features, the perceptual and motor demands of a task, were historically assumed to preserve orthogonality, on the premise that decision processes are fully separable from sensorimotor ones. This premise is trivially true under the classical serial model of brain function, in which sensory stimuli are first encoded into abstract representations, decision processes then operate over these representations to produce a choice, and the motor system finally implements it \cite{sternberg_discovery_1969}. A growing body of evidence, however, suggests this strict segregation does not hold \cite{gallivan_decision-making_2018, cisek_neural_2010}. The visual salience of an option can inflate its apparent value, biasing choice toward whichever option stands out most \cite{milosavljevic_relative_2012}. The motor cost of acting, such as the amount of turning needed to reach an option, can similarly shift which option is chosen \cite{griesbach_embodied_2022}. These effects need not even run through value: when the same person makes the same choice via joystick, manual reach, or eye movement, the rate of evidence accumulation changes \cite{ivanov_decision-making_2024}. Together, these studies point to several distinct pathways through which the sensorimotor components of a task reach into the decision process, underscoring the risk that sensorimotor variability introduces the non-orthogonal variance described above.

Sensorimotor features already vary across standard computational psychiatry paradigms. In MRI studies, participants lie supine in a loud, confined space with restricted movement and respond using a button box; in purely behavioral tasks, participants sit upright at a desk and respond with a keyboard or mouse. This difference is not trivial, as supine posture alone delays reaction times on cognitive tasks relative to upright posture \cite{muehlhan_effect_2014, sun_supine_2021}. The field's current trajectory will only increase this heterogeneity. Ecological momentary assessment, a rapidly growing approach in psychiatric research, administers tasks via smartphone during daily life, so the same participant may complete a task while lying in bed, commuting, or walking \cite{russell_annual_2020}. Virtual reality tasks, increasingly explored for psychiatric assessment, introduce entirely different visual environments and movement requirements relative to standard button-press tasks \cite{geraets_use_2022}. The cumulative effect is a field in which sensorimotor features shift substantially across paradigms, motivating the need to investigate how such shifts introduce systematic nuisance variance into tasks and measures.

Beyond concurrent effects, the sensorimotor features of a participant's recent past are also important to consider. In proposed clinical deployment, computational psychiatry tasks would be administered in batteries, meaning each task would be preceded by another with potentially different sensorimotor demands \cite{benrimoh_barriers_2023}. Participants also arrive with a variable sensorimotor history, having been sedentary, physically active, or anywhere in between in the hours before testing. Prior sensorimotor experience thus represents a second channel through which measurement validity could be compromised.

To examine if and how variations in sensorimotor features bias decision behavior, we designed a two-choice reversal learning task that participants completed under two different sensorimotor conditions (Figure 1). In the stationary condition, participants needed only arm movements to make choices; in the active condition, participants needed to walk between physically separated computers to make choices. The task structure was the same across both conditions: participants chose one of two options and either received a reward (1 point) or no reward. The reward contingency was set at 70$\%$ for one option and 30$\%$ for the other, with the contingencies periodically reversing. This reversal learning structure was chosen because it is among the most widely used tests of cognitive flexibility across neuropsychiatric conditions \cite{izquierdo_neural_2017}, and because the reversals require continual tracking of the current task contingency. Participants completed both conditions, with random assignment of starting condition, allowing investigation of order effects, specifically whether prior sensorimotor experience shapes subsequent decision behavior. We quantified behavior using three common measures: accuracy, stay rate, and weights from trial-by-trial logistic regressions predicting the next choice from recent choice-reward history. No prior work has directly manipulated sensorimotor features within a reversal learning paradigm, so we had no basis to predict which of these measures, if any, would prove sensitive to condition.

\begin{figure}[hbt!]
    \centering
    \includegraphics[width=1\linewidth]{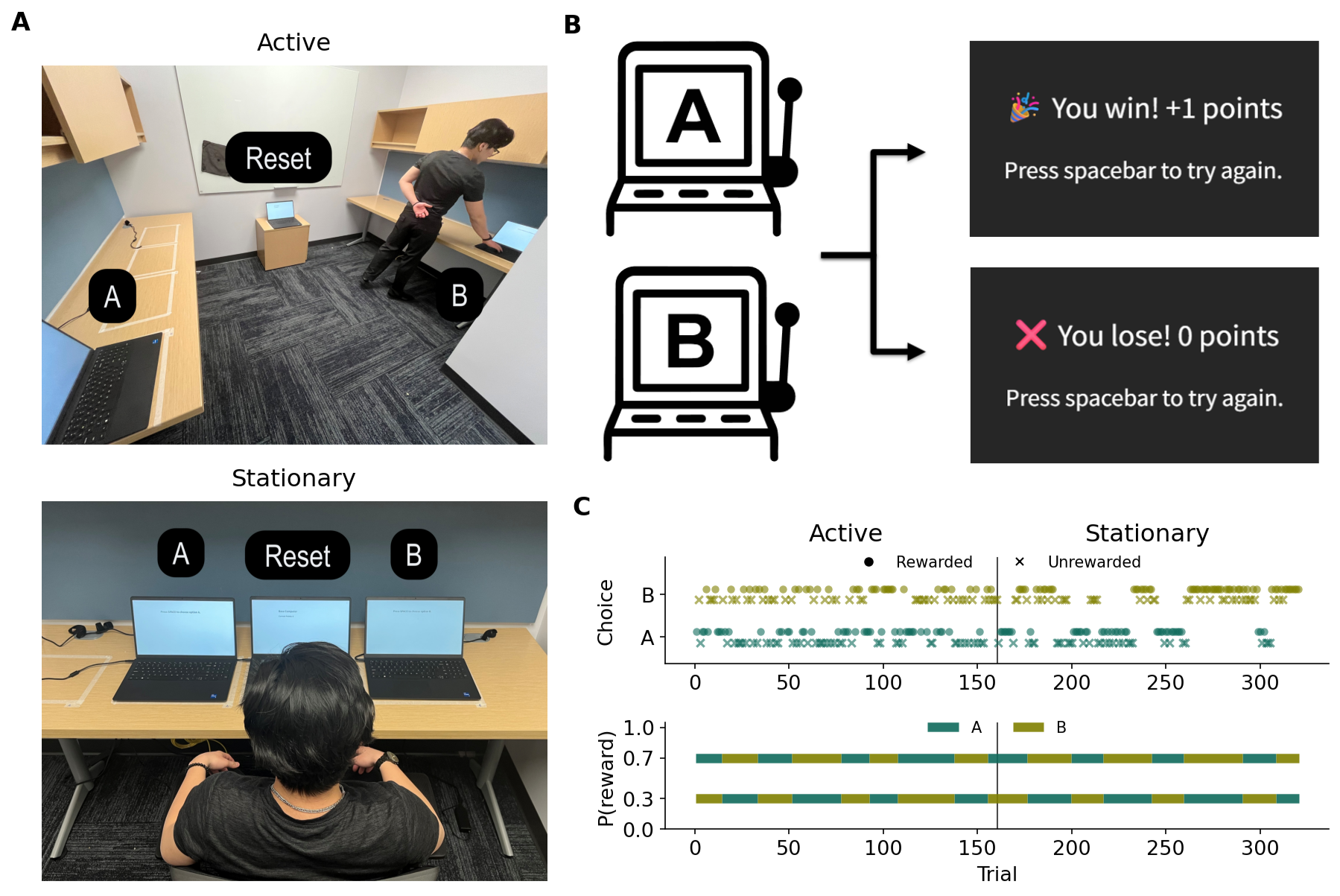}
    \caption{\textbf{Task design and structure.} (A) Participants completed a two-choice reversal learning task under two sensorimotor conditions. In the Active condition (top), physically separated laptops (labeled A and B) required participants to walk to the option computer to make the choice, and then had to walk back to the central "Reset" station to start the next trial. In the Stationary condition (bottom), the same A/B choice and Reset options were presented on adjacent laptops within arm's reach, requiring only arm movements. The individual pictured is an author of this study and has given consent for the image to be published. (B) On each trial, participants selected option A or B by pressing the spacebar on that option's respective computer and received feedback indicating a win (+1 point) or loss (0 points). Participants then pressed the spacebar on the "Reset" computer to begin the next trial. (C) Example session from a participant in the active-first order. The vertical black line marks the transition between conditions (trials 1–160: Active; trials 161–320: Stationary) and applies to both plots below. Top: choice sequence across trials, with option (A: teal, B: olive) and trial outcome (rewarded: circle; unrewarded: x) shown for each trial. Bottom: reward contingency schedule for the same session. Contingencies (70$\%$/30$\%$, alternating between A and B) reversed after a minimum of 15 trials with a 20$\%$ probability per trial, yielding a variable inter-reversal interval with an expected length of 20 trials.}
    \label{fig:placeholder}
\end{figure}
\FloatBarrier 

\section*{Results}

\subsection*{Participants and task factors}

Participants were 90 individuals recruited from a university research participation pool who received course credit for participation. Three factors are referenced throughout: previous outcome (win vs. lose on the prior trial), a within-subject factor reflecting outcome-dependent choice behavior; condition (active vs. stationary), the within-subject manipulation of sensorimotor features; and order (active-first vs. stationary-first), a between-subject factor reflecting which starting condition participants were assigned to. Additionally, a fourth factor, report (report vs. no-report), reflects the latter half of participants recruited, who completed an open-ended prompt describing their decision strategy in addition to the task. Report is included to verify that this procedural addition did not significantly alter behavior, but is not otherwise a factor of interest for this paper.

\subsection*{Task performance does not vary with sensorimotor features}

Participants performed the task in a reward-driven manner. During stable periods (trials 8 or more after a reversal), choice of the optimal option (the 70\% reward option) was above chance (t(89) = 11.95, p < .001, d = 1.26) and returned to this level following each reversal (Figure 2), confirming that choices tracked the prevailing contingency and reorganized when it changed. 

A condition (within) x order x report (between) repeated-measures ANOVA (Supplementary Table 1) on stable-period accuracy revealed no significant main effects or two-way interactions. A three-way order x report x condition interaction reached significance (F(1, 86) = 4.33, p = .040, $\eta^2_G$ = .014), but Holm-corrected post-hoc contrasts were all non-significant. This suggests that performance is equivalent across conditions. 

\begin{figure}[hbt!]
    \centering
    \includegraphics[width=1\linewidth]{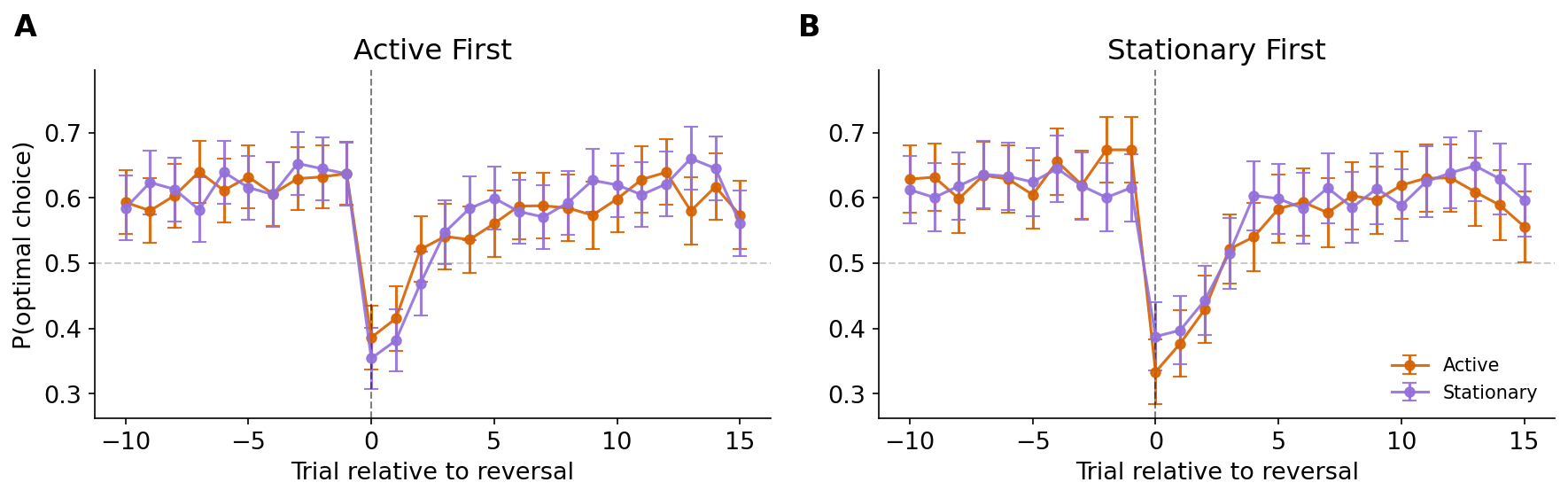}
    \caption{\textbf{Reversal learning curves.} Split by order (A: Active First, B: Stationary First) and colored by condition (Orange: Active, Purple: Stationary). Mean P(optimal choice) ± 95$\%$ CI is shown across trials aligned to each reward contingency reversal (dashed vertical line at trial 0). Horizontal dashed line indicates chance (0.5). Asterisks denote trials where Active and Stationary accuracy differed significantly (paired t-tests, Benjamini-Hochberg FDR-corrected: * p < .05, ** p < .01, *** p < .001).}
    \label{fig:2}
\end{figure}

\subsection*{Sensorimotor features have an order-dependent effect on stay rate}

Stay rate is the proportion of trials on which the participant repeated their previous choice. A previous outcome x condition (within) x order x report (between) mixed ANOVA (Supplementary Table 2) on stay rate revealed a significant main effect of previous outcome (F(1, 86) = 153.65, p < .001, $\eta^2_G$ = .344), confirming expected win-stay/lose-shift behavior.

\begin{figure}[hbt!]
    \centering
    \includegraphics[width=1\linewidth]{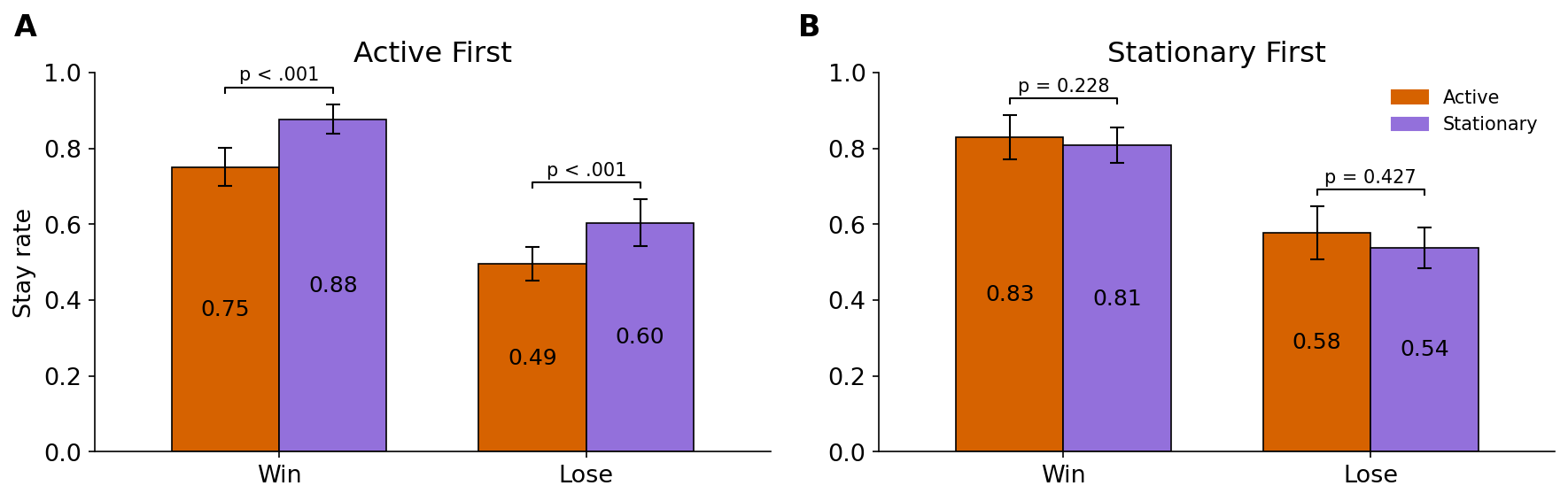}
    \caption{\textbf{Win-stay and lose-stay rate bar plots.} Split by order (A: Active First, B: Stationary First) and colored by condition (Orange: Active, Purple: Stationary). Displays mean stay rate ± 95$\%$ CI divided by previous outcome (Win vs. Lose). Brackets indicate Holm-corrected post-hoc t-tests between Active and Stationary conditions, following a repeated-measures ANOVA. }
    \label{fig:placeholder}
\end{figure}

A significant main effect of condition was also observed (F(1, 86) = 11.56, p = .001, $\eta^2_G$ = .016), though this effect was qualified by a significant two-way interaction between order and condition (F(1, 86) = 31.54, p < .001, $\eta^2_G$ = .038). The interaction explained more than twice the variance of the main effect. Holm-corrected post-hoc t-tests (Figure 3) indicated this interaction was driven by participants who completed the active condition first, who showed significantly lower stay rate in the active condition relative to stationary (t(86) = -6.81, p < .001). The comparison for those who completed the stationary condition first was not significant (t(86) = 1.38, p = .171). Given that the condition effect was present only in the active-first group, the significant condition main effect is better understood as a consequence of this interaction than as evidence of a general effect of condition independent of order.

A significant three-way interaction between report, condition, and previous outcome was observed (F(1, 86) = 4.25, p = .042, $\eta^2_G$ = .005), as was a significant four-way interaction among order, report, condition, and previous outcome (F(1, 86) = 3.96, p = .047, $\eta^2_G$ = .005); both effect sizes were negligible and are not interpreted further. Given the minute interactions and the report factor's lack of theoretical relevance to the sensorimotor manipulation, it is dropped from subsequent analyses.

\subsection*{Individual differences in stay rate are stable, but shifted between order groups}

Stay rate correlated strongly across active and stationary conditions in both order groups for both win-stay (Active First: r(46) = .68, p < .001; Stationary First: r(40) = .59, p < .001) and lose-stay (Active First: r(46) = .48, p = .001; Stationary First: r(40) = .67, p < .001), indicating stable individual differences across sensorimotor conditions (Figure 4). 

Regression of stationary on active stay probability revealed no significant order x slope interaction for win-stay (p = .656) or lose-stay (p = .396), but significant displacements of +0.11 for win-stay (p < .001) and +0.11 for lose-stay (p = .002). This indicates that the order-dependent effect of sensorimotor features reflects a uniform displacement of individual baselines.

\begin{figure} [hbt!]
    \centering
    \includegraphics[width=1\linewidth]{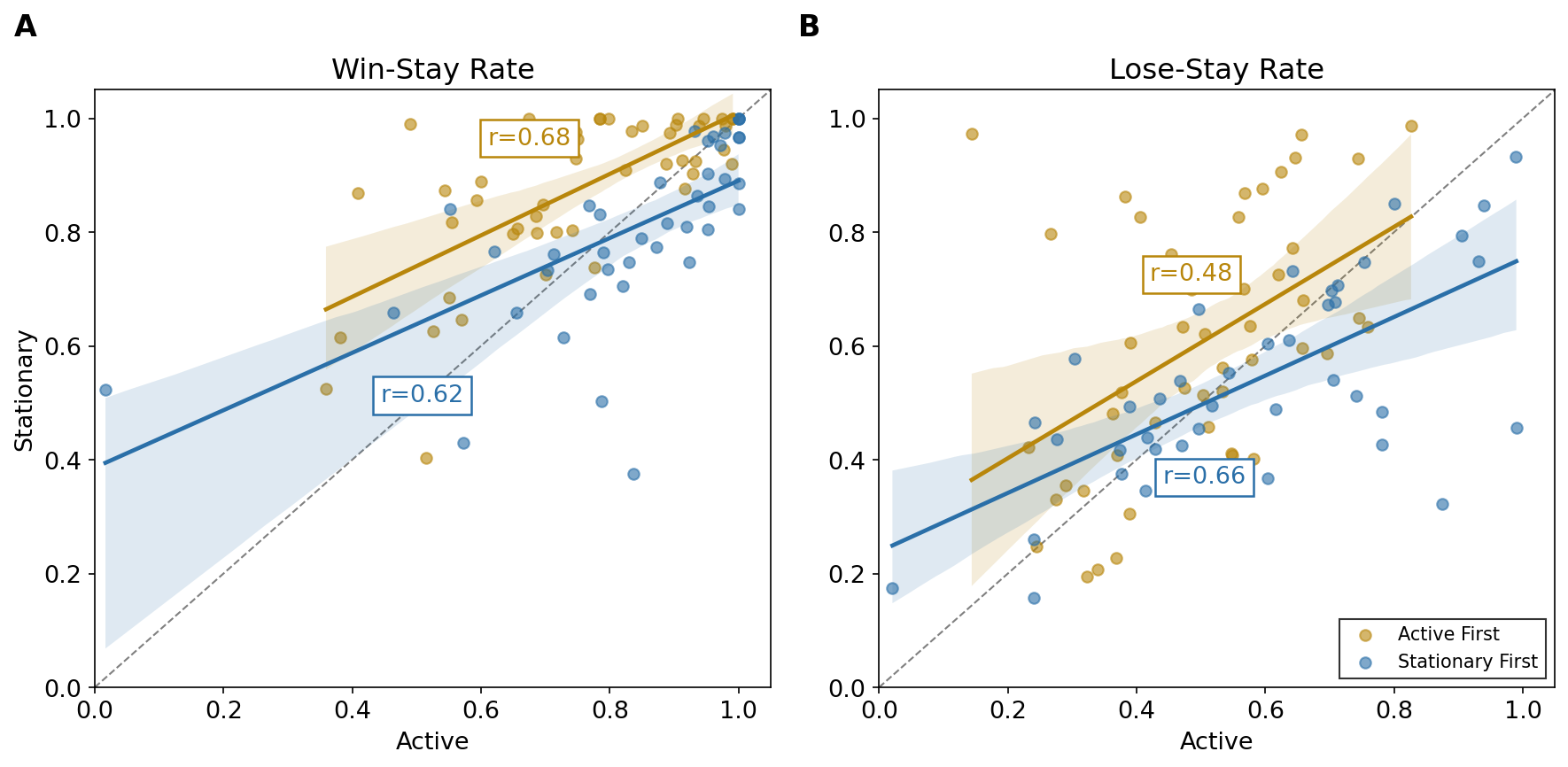}
    \caption{\textbf{Win-stay and lose-stay rate correlations.} Split by previous outcome (A: Win, B: Lose) and colored by order (Yellow: Active First, Blue: Stationary First). Each point represents one participant's stay rate in the Stationary condition plotted against the Active condition. Lines show order-specific linear fits with 95$\%$ CI bands; the dashed diagonal indicates unity (x = y).}
    \label{fig:placeholder} 
\end{figure}

\subsection*{The active-to-stationary transition shifts win-history weighting}

We fit per-subject logistic regressions predicting current choice from win and lose history over the preceding five trials, estimating separate weights for each condition and t-testing condition differences lag-by-lag with FDR correction for each group-condition combination. One-sample t-tests revealed that win weights were significantly above zero at lag 1 across both order groups and both conditions, and at lag 2 in three of the four group-condition combinations, with no consistent influence at lags 3 to 5 (Supplementary Table 3). This suggests that previous wins influenced choices beyond the immediately succeeding trial.

\begin{figure}[hbt!]
    \centering
    \includegraphics[width=1\linewidth]{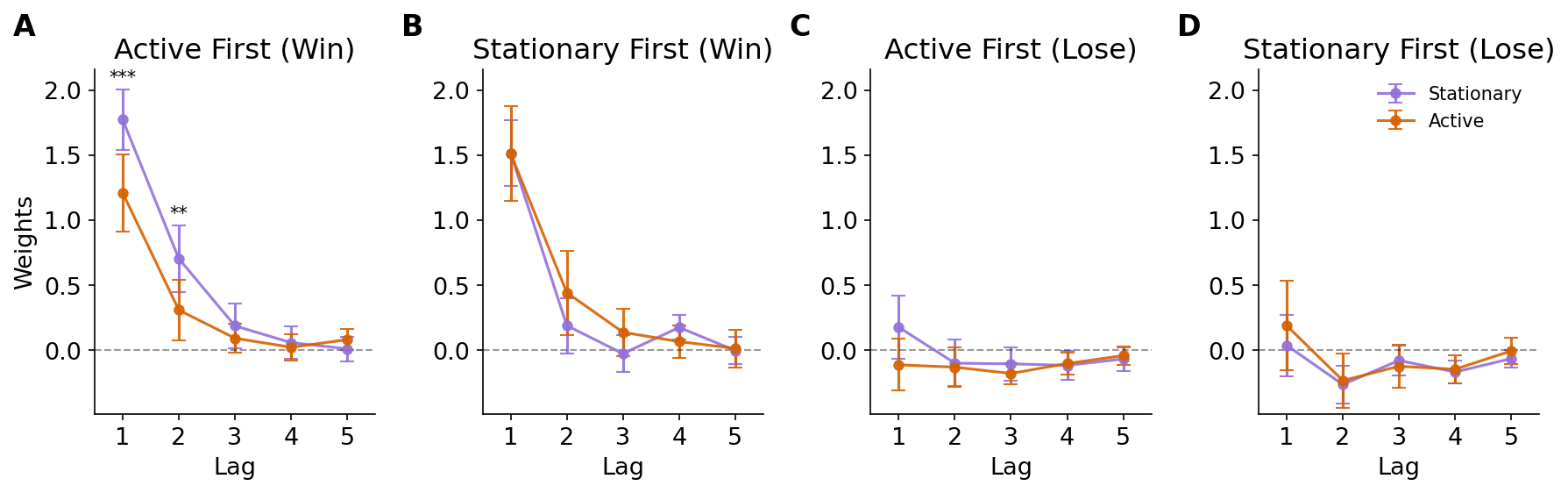}
    \caption{\textbf{Win and lose history weights.} Split by previous outcome and order (Win—A: Active First, B: Stationary First; Lose—C: Active First, D: Stationary First) and colored by condition (Orange: Active, Purple: Stationary). Shows mean logistic regression coefficients ± 95$\%$ CI across lags 1–5. Dashed horizontal line indicates zero. Asterisks denote lags where Active and Stationary weights differed significantly (paired t-tests, Benjamini-Hochberg FDR-corrected: * p < .05, ** p < .01, *** p < .001).}
    \label{fig:placeholder}
\end{figure}

Pairwise t-tests between conditions (Supplementary Table 4) showed higher win weights for the stationary block in the Active First group at both lag 1 (t(47) = -4.49, p < .001, d = 0.59) and lag 2 (t(47) = -3.69, p = .001, d = 0.45), with no significant difference at lags 3 to 5. In the Stationary First group, no condition contrast survived FDR correction at any lag. The selective elevation of win-history weights in the Active First group mirrors the order-dependent pattern in stay rate, demonstrating that history weighting is similarly modified by the active-to-stationary transition.

Lose weights were an order of magnitude smaller than win coefficients and no condition contrast survived FDR correction (smallest p = .090); lose-driven switching is not interpreted further.

\subsection*{Individual differences maintain stability across win-history weights}

Individual differences in win weighting were preserved across conditions in both order groups at both lag 1 (Active First: r(46) = .60, p < .001; Stationary First: r(40) = .58, p < .001) and lag 2 (Active First: r(46) = .65, p < .001; Stationary First: r(40) = .65, p < .001), indicating a stable individual differences structure across sensorimotor conditions (Figure 6).

\begin{figure}[hbt!]
    \centering
    \includegraphics[width=1\linewidth]{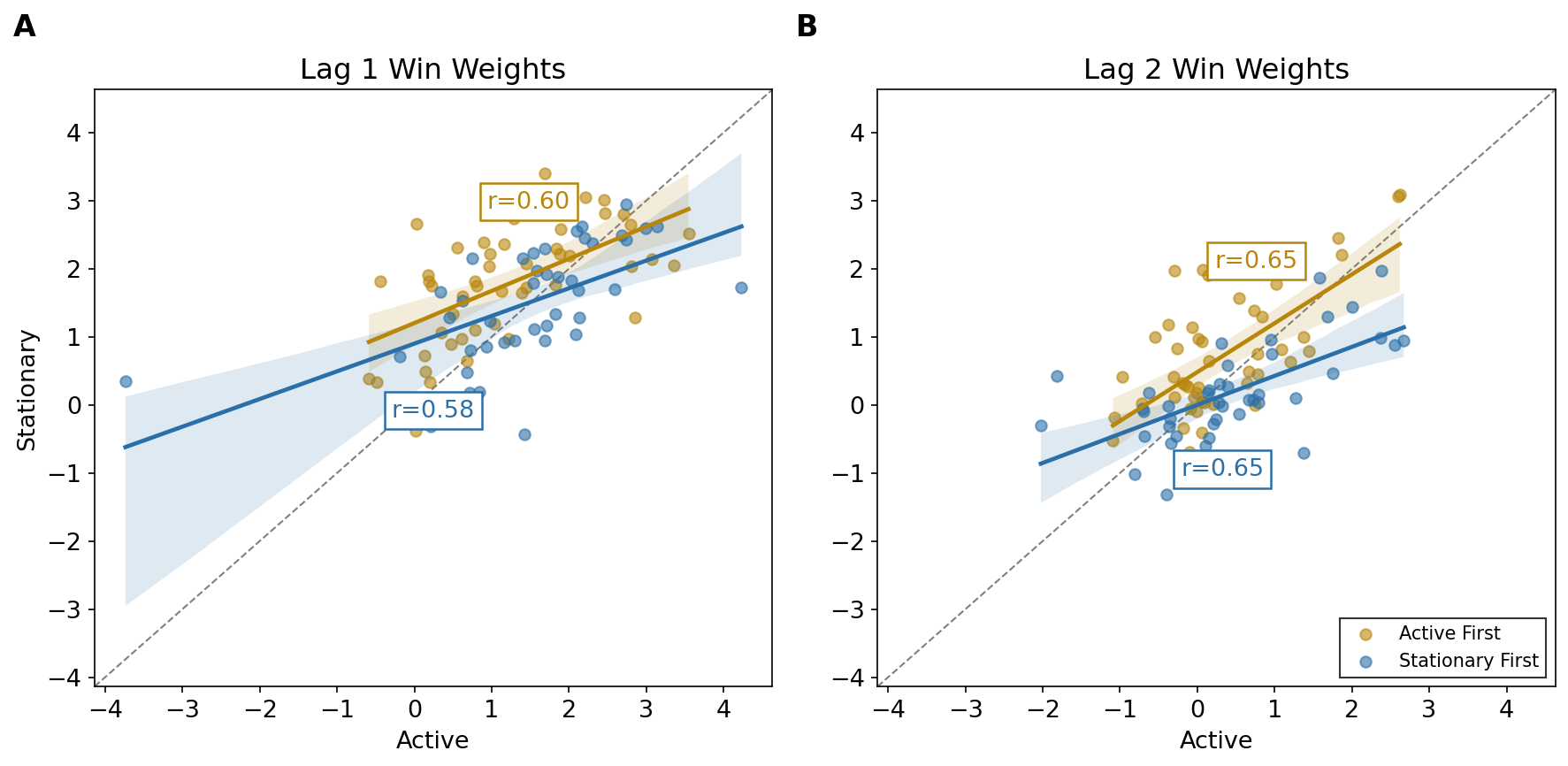}
    \caption{\textbf{Lag 1 and lag 2 win weight correlations.} Split by lag (A: lag 1, B: lag 2) and colored by order (Yellow: Active First, Blue: Stationary First). Each point represents one participant's stationary-condition win weight plotted against their active-condition win weight. Lines show order-specific linear fits with 95$\%$ CI bands; the dashed diagonal indicates unity (x = y).}
    \label{fig:placeholder}
\end{figure}

\FloatBarrier 

The order-dependent sensorimotor effect differs by lag. At lag 1, regression of stationary on active win weights revealed no significant order x slope interaction (p = .614) and a significant mean displacement of +0.39 (p = .009), indicating a uniform shift of individual baselines consistent with the stay-rate pattern. At lag 2, the order x slope interaction was significant (p = .048), with a steeper slope in the Active First group (0.72) than the Stationary First group (0.43). The mean displacement was also larger at +0.59 (p < .001). This suggests that the effect acts heterogeneously across individual baselines at lag 2.

\section*{Discussion}

In this study, we examined whether changing the sensorimotor features under which a reversal learning task is completed biases decision behavior. Our results suggest that sensorimotor variations do affect the behavioral measures computed, but mainly as a function of which condition came first. Specifically, participants were more likely to repeat their previous choice in the stationary condition when it followed the active condition, whereas when the stationary condition came first, stay rate did not differ significantly between conditions. A logistic regression replicated and extended this order-dependent effect, showing that the increase reflects greater weighting of past wins in current decision-making. This shift occurred without any corresponding difference in accuracy, indicating that it is not a simple alteration of task engagement but a more targeted modulation of the decision-making process. Cross-condition correlations in stay rate and win-history weighting were strong and consistent across order groups and behavioral measures, suggesting that the sensorimotor manipulation does not disrupt individual differences but does cause a systematic increase in their absolute values.

The implications are mixed. Positively, this suggests the task and quantitative measures index a stable component of cognition that differs between participants: those with high values in one sensorimotor condition tended to show high values in the other, and low values tracked low values across conditions. The absolute values are shifted between conditions, but this offset appears correctable, in theory, by identifying and canceling the offset and slope multiplier between conditions. If so, the resulting stable individual differences have potential value for psychiatric phenotyping.

How remediable the order-dependent effect is, however, depends on how broadly it generalizes across two dimensions. The first is breadth: if the bias transfers to other tasks with different sensorimotor features, the composition and ordering of a clinical assessment battery could confound latent variable estimates unless standardized or explicitly modeled. The second is duration: if the effect persists across longer time windows, the sensorimotor features of activities participants engage in hours before assessment could shift their choices. Because it is practically impossible to estimate the exact offset and multiplier for such sources of variance, this is a potentially consequential threat to proposed clinical applications. If the behavioral measures are used as clinical decision values, the same person could be evaluated as above or below a clinical threshold depending on their prior sensorimotor history. The present study cannot determine where on either dimension the effect sits, and therefore cannot bound the severity of the threat it identifies. This motivates future work clarifying the generalizability of the findings.

A further limitation is that the present design cannot isolate which aspect of the sensorimotor manipulation drives the order-dependent shift. The active and stationary conditions differed along multiple dimensions simultaneously, including physiological arousal, effort costs, and the nature of sensorimotor engagement during choice. Future work should hold all but one of these constant to identify the active ingredient. A final limitation is that the sample was drawn from a healthy university population. It remains unclear how psychopathology might interact with the mechanism underlying this effect, potentially modulating its magnitude, direction, or even presence. Testing in clinical populations is an important next step, since these are the populations computational psychiatry aims to better characterize.

The sensorimotor effect identified here is one instance of a broader class of factors that can violate the orthogonality assumption underlying the cognitive-task approach of computational psychiatry. Stress decreases learning rates and increases reliance on habitual behavior \cite{park_stress_2017, carvalheiro_acute_2021}. Greater cognitive load increases lose-shift choices \cite{ivan_lose-shift_2018} and promotes exploitative behavior \cite{brown_humans_2022}. Arousal state shapes the character of decision-making, with low arousal producing greater decision volatility and high arousal increasing perseverative behavior \cite{ciria_different_2021}. Attentional engagement further modulates valuation, with greater attention causing appetitive options to be chosen more frequently and aversive options less so \cite{armel_biasing_2008}. Most similarly to the present study, greater physical effort expended in making choices increases learning rates following positive outcomes and decreases them following negative outcomes \cite{jarvis_effort_2022}. These represent only a subset of the variables present for any individual completing a cognitive task. What unifies them is the risk they pose to measurement: each can introduce systematic, directionally structured variance into choices. This underscores the need to investigate how many such contaminating variables exist, how large their effects are, and how assessment practices can mitigate their effects. Such validation efforts are necessary for the cognitive-task approach to achieve clinical utility.

\section*{Methods}

\subsection*{Participants}

Ninety-six individuals were recruited from a university research participation pool and received course credit for participation. Six were excluded prior to analysis: four for behavioral exclusion criteria (win-stay rate below 0.5 in both conditions, or an invariant response strategy defined as stay rate > .95 or < .05 in both conditions), one for failing to perform the task correctly, and one for data loss. The final sample comprised 90 participants (demographic data available for n = 87: M age = 19.5, SD = 1.6, range = 18-29; 50 men, 36 women, 1 other). Participants were randomly assigned to complete the active condition first (n = 48) or the stationary condition first (n = 42). A subset of participants (n = 48) additionally completed open-ended strategy reports following each condition (see Procedure). All procedures were approved by the Georgia Institute of Technology Institutional Review Board ($\#$IRB2025-388), and informed consent was obtained from all participants.

\subsection*{Task}

The task was a two-choice probabilistic reversal learning bandit task performed under two sensorimotor conditions (Stationary and Active). One option yielded a reward (1 point) with 70$\%$ probability (advantageous); the other yielded a reward with 30$\%$ probability (disadvantageous). After a minimum of 15 trials, the identity of the advantageous option reversed with a 20$\%$ probability on each subsequent trial, requiring participants to continually track which option was currently advantageous to maintain task performance. Participants were given no information about the task's probabilistic structure or the possibility of reversals.

On each trial, participants pressed the spacebar at a base computer, moved to their chosen option, and pressed the spacebar on the option computer to confirm their choice, before returning to the base computer. In the stationary condition, all computers were arranged in parallel within arm's reach, and choices required only arm movements. In the active condition, computers were placed on separate tables, requiring participants to walk between computers to make each choice. This manipulation held the task structure constant across conditions while varying the sensorimotor demands of making a choice.

\subsection*{Apparatus}

Data collection used three identical Dell 15 laptops running Windows 11 Pro. One laptop connected to the university's eduroam network and hosted a local Wi-Fi hotspot; the remaining two laptops connected to this hotspot to receive synchronized task stimuli. The task was custom-coded primarily using the Streamlit package.

\subsection*{Procedures}

Upon arrival, participants provided informed consent and received standardized on-screen instructions. Participants then completed five practice trials to familiarize themselves with the task mechanics. Immediately following practice, baseline affect was assessed via continuous 0-10 ratings of valence and arousal.

Participants completed 160 trials per condition (320 trials total), divided into two 80-trial blocks per condition. The task was self-paced, and feedback was delivered on-screen after each trial ("You win! +1 points" or "You lose! 0 points"). Following each block, participants also completed continuous 0-10 ratings of affect (valence, arousal) and task experience (mental load, physical load, sense of success, sense of control, attention). Following the second block of the first condition, participants took a 5-minute break before beginning the second condition; during this break, participants also completed a brief survey collecting demographic information and potentially behaviorally relevant state variables such as sleep quality.

Partway through data collection, an additional measure was introduced: following the final block of each condition, participants provided a typed open-ended description of their decision-making approach. Because this measure was added after data collection had begun, it was completed only by the subset of participants enrolled from that point onward (n = 48).

Upon session completion, participants were debriefed on the task's true reward contingencies and awarded course credit. Analysis and interpretation of the subjective ratings, strategy reports, and state variables (aside from basic demographics) are beyond the scope of the present paper.

\subsection*{Accuracy}

Accuracy was calculated as the proportion of trials on which the participant selected the advantageous option. Because the identity of the advantageous option reversed periodically, accuracy is necessarily lower in the trials immediately following a reversal, before participants have had the opportunity to detect the change and shift their choice accordingly. To obtain a stable measure of performance, accuracy was therefore calculated using only trials occurring at least 8 trials after the most recent reversal, separately for each participant in the active and stationary conditions.

We first tested whether overall accuracy exceeded chance (50$\%$) using a one-sample t-test. We then tested whether accuracy differed between conditions using a mixed ANOVA with condition (active vs. stationary) as a within-subject factor and order (active-first vs. stationary-first) and report (report vs. no report) as between-subject factors. Effect sizes are reported as generalized eta-squared, $\eta^2_G$. Significant interactions were followed up with pairwise post-hoc comparisons, Holm-corrected for multiple comparisons.

To examine performance at the resolution of individual trials, we computed the trial-level accuracy for every participant and condition at each trial relative to the nearest reversal, spanning from 10 trials before to 15 trials after the reversal. Active and stationary conditions were compared using paired t-tests at each relative trial position, performed separately within each order group, with FDR correction applied across trial positions.

\subsection*{Stay Rate}

Stay rate was calculated as the proportion of trials on which a participant repeated their choice from the immediately preceding trial. Trials at the start of a block, for which the preceding trial belonged to the other condition, were excluded, since the previous choice and outcome could not be attributed to the current condition in these cases. Stay rate was calculated separately for each participant, condition (active, stationary), and previous outcome (win-stay: following a rewarded trial; lose-stay: following an unrewarded trial).

We tested whether stay rate differed across conditions and previous outcomes using a mixed ANOVA with condition (active vs. stationary) and previous outcome (win vs. lose) as within-subject factors, and order (active-first vs. stationary-first) and report (report vs. no report) as between-subject factors. Effect sizes are reported as generalized eta-squared, $\eta^2_G$. Significant interactions were followed up with pairwise post-hoc comparisons, Holm-corrected for multiple comparisons.

\subsection*{History Regression}

To characterize how recent reward history influenced choice, we fit a logistic regression predicting each trial's choice from the outcomes of the preceding five trials (lags 1-5), separately for each participant and condition. Choice was coded as whether the participant chose option A on the current trial (A = 1, B = 0). At each lag, win predictors were coded $+$1 if the choice at that lag was option A and rewarded, $-$1 if option B and rewarded, and 0 otherwise. Lose predictors were coded analogously: $+$1/$-$1 for option A/B chosen on an unrewarded trial, and 0 otherwise. Trials for which any of the preceding five trials belonged to the other condition were excluded, so that history effects reflected only within-condition trial sequences. Separate logistic regressions were fit using win predictors and lose predictors, yielding a win-history model and a lose-history model, each comprising five weights (lags 1-5).

For each order group (active-first, stationary-first), the group-level analysis was restricted to participants with valid coefficients in both conditions. At each lag, we tested whether win- and lose-history weights differed from zero using one-sample t-tests, and whether they differed between the active and stationary conditions using paired t-tests, with FDR correction applied across lags separately for each test family.

To confirm that this regression procedure could reliably recover the history weights, we conducted a parameter recovery analysis. For each participant and condition, the fitted win- and lose-history weights (five lags each) served as generative parameters for a synthetic agent performing a two-armed bandit task with reversals. Each parameter set (participant x condition) was used to generate 100 synthetic agents, each completing 160 trials. For every synthetic dataset, the same win-history and lose-history logistic regressions used in the empirical analysis were refit to recover the generating weights; recovered weights were then averaged across the 100 simulations per participant-condition. Averaged recovered weights were compared to the true, fitted weights that generated them using Pearson correlation, computed separately for each lag and for the win- and lose-history models. Recovered weights correlated strongly with the true weights (r = 0.81-0.97, all p < .001) across all lags (Figure 7), confirming that the regression procedure reliably recovers history weights from behavior at this trial length.

\begin{figure}[hbt!]
    \centering
    \includegraphics[width=1\linewidth]{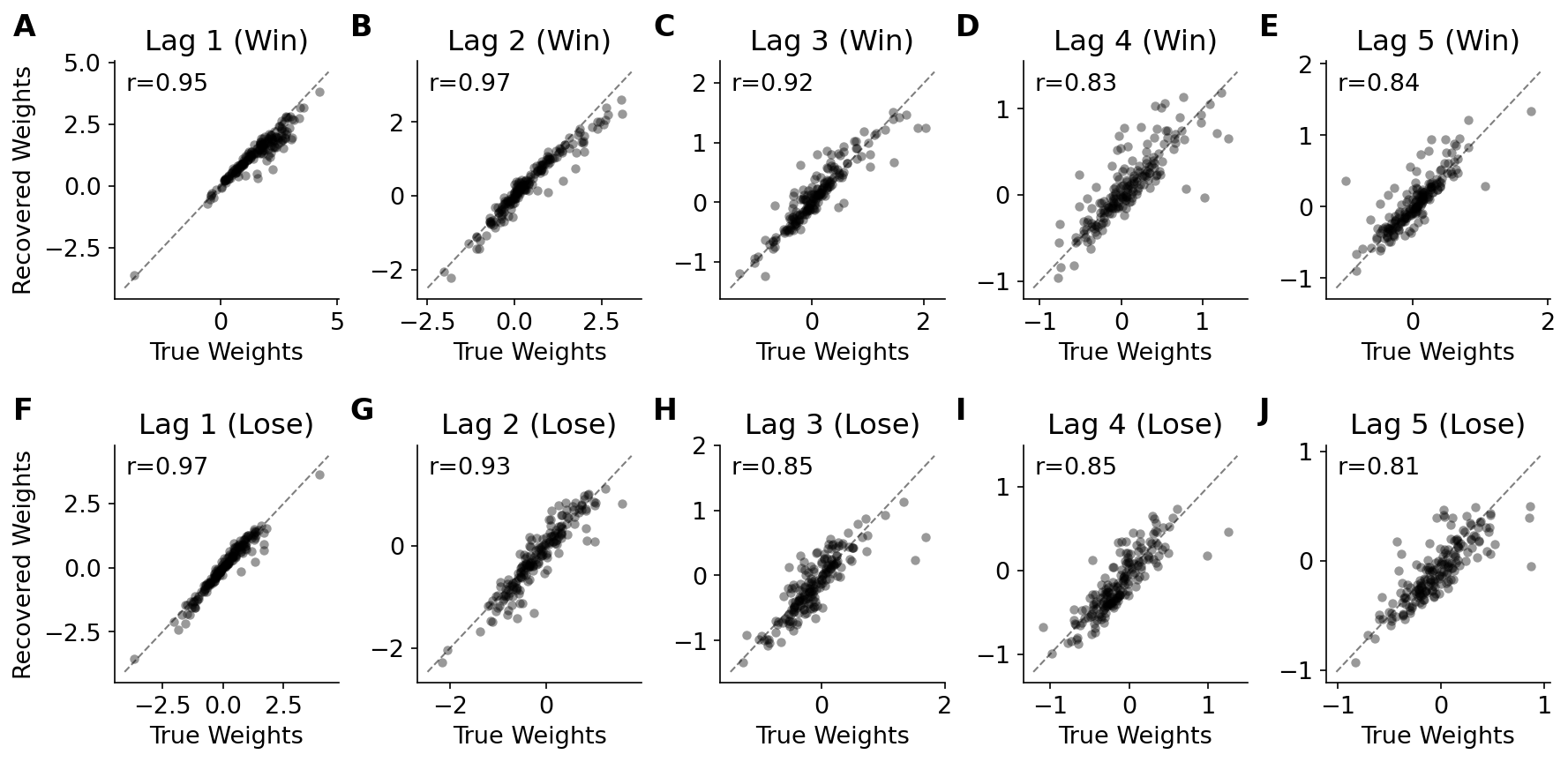}
    \caption{\textbf{Recovery of history regression weights.} Recovered versus true win-history (A–E) and lose-history (F–J) weights split across lags 1–5. Each point represents one participant-condition's fitted (true) weight plotted against the weight recovered by refitting the win- or lose-history logistic regression to synthetic choice data generated using that weight, averaged across 100 simulated agents. The dashed diagonal indicates unity (x = y).}
    \label{fig:placeholder}
\end{figure}

\FloatBarrier 

\subsection*{Individual Differences}

To assess whether participants who showed a given pattern of choice behavior in one condition also showed it in the other, we examined the correlation between active condition and stationary condition values at the level of individual participants. This was performed for two sets of measures: win-stay and lose-stay rates (see Stay Rate), and win-history regression weights at lags 1 and 2 (see History Regression). For each participant, trials were averaged within condition to yield a single active condition and stationary condition value per measure. The active-stationary relationship was assessed separately within each order group (active-first, stationary-first) using Pearson correlation.

To test formally whether the active-stationary relationship differed between order groups, we fit a linear regression predicting stationary-condition values from mean-centered active-condition values, order, and their interaction. A significant interaction indicated that the active-stationary relationship (i.e., the slope) differed between order groups. A significant coefficient for order indicated a mean shift between order groups.

\subsection*{Software}

Mixed ANOVAs and post-hoc pairwise comparisons were conducted in R using the ez and emmeans packages. Correlation and t-tests were performed in Python using pingouin; the active/stationary order-interaction models used statsmodels. Trial-history logistic regression models of choice behavior, including the model recovery simulations, were fit using scikit-learn.

\bibliography{references}

\section*{Acknowledgements}

We acknowledge the use of Claude (Anthropic) for coding assistance and editorial refinement during the preparation of this manuscript.

\section*{Author contributions statement}

E.H. and R.W. conceived the experiment, E.H. and W.X. developed the task software, E.H. and W.X. conducted the experiment, and E.H. analysed the results.  All authors reviewed the manuscript. 

\section*{Additional information}

Supplementary Information accompanies this paper. Code and data are available on Zenodo at https://doi.org/10.5281/zenodo.21815154.

\clearpage
\renewcommand{\thetable}{S\arabic{table}}
\setcounter{table}{0}

\section*{Supplementary Information}

\begin{table}[ht]
\centering
\begin{tabular}{lrrrrr}
\hline
Effect & $F$ & $df_1$ & $df_2$ & $p$ & $\eta^2_G$ \\
\hline
Order & 0.00 & 1 & 86 & .954 & .000 \\
Report & 0.11 & 1 & 86 & .739 & .001 \\
Condition & 0.17 & 1 & 86 & .680 & .001 \\
Order x Report & 0.93 & 1 & 86 & .339 & .008 \\
Order x Condition & 2.01 & 1 & 86 & .160 & .006 \\
Report x Condition & 0.03 & 1 & 86 & .859 & .000 \\
Order x Report x Condition & 4.33 & 1 & 86 & .040 & .014 \\
\hline
\end{tabular}
\caption{Condition (within) x order x report (between) mixed ANOVA on stable-period accuracy.}
\end{table}

\begin{table}[ht]
\centering
\begin{tabular}{lrrrrr}
\hline
Effect & $F$ & $df_1$ & $df_2$ & $p$ & $\eta^2_G$ \\
\hline
Order & 0.02 & 1 & 86 & .893 & .000 \\
Report & 1.77 & 1 & 86 & .187 & .010 \\
Condition & 12.55 & 1 & 86 & .001 & .016 \\
Previous Outcome & 153.59 & 1 & 86 & $<$.001 & .344 \\
Order x Report & 1.54 & 1 & 86 & .219 & .009 \\
Order x Condition & 31.26 & 1 & 86 & $<$.001 & .038 \\
Report x Condition & 0.00 & 1 & 86 & .956 & .000 \\
Order x Previous Outcome & 0.03 & 1 & 86 & .861 & .000 \\
Report x Previous Outcome & 1.73 & 1 & 86 & .192 & .006 \\
Condition x Previous Outcome & 0.02 & 1 & 86 & .902 & .000 \\
Order x Report x Condition & 0.95 & 1 & 86 & .334 & .001 \\
Order x Report x Previous Outcome & 0.07 & 1 & 86 & .799 & .000 \\
Order x Condition x Previous Outcome & 0.33 & 1 & 86 & .570 & .000 \\
Report x Condition x Previous Outcome & 4.25 & 1 & 86 & .042 & .005 \\
Order x Report x Condition x Previous Outcome & 4.06 & 1 & 86 & .047 & .005 \\
\hline
\end{tabular}
\caption{Previous outcome x condition (within) x order x report (between) mixed ANOVA on stay rate.}
\end{table}

\begin{table}[ht]
\centering
\begin{tabular}{llrrrrr}
\hline
Order & Condition & Lag & $t$ & $df$ & $p_{\text{FDR}}$ & $d$ \\
\hline
Active First & Active & 1 & 7.95 & 47 & $<$.001 & 1.15 \\
             &        & 2 & 2.57 & 47 & .034 & 0.37 \\
             &        & 3 & 1.57 & 47 & .155 & 0.23 \\
             &        & 4 & 0.41 & 47 & .682 & 0.06 \\
             &        & 5 & 1.94 & 47 & .098 & 0.28 \\
\hline
Active First & Stationary & 1 & 14.77 & 47 & $<$.001 & 2.13 \\
             &            & 2 & 5.28 & 47 & $<$.001 & 0.76 \\
             &            & 3 & 2.09 & 47 & .070 & 0.30 \\
             &            & 4 & 0.90 & 47 & .467 & 0.13 \\
             &            & 5 & 0.20 & 47 & .841 & 0.03 \\
\hline
Stationary First & Active & 1 & 8.07 & 41 & $<$.001 & 1.24 \\
                 &        & 2 & 2.63 & 41 & .030 & 0.41 \\
                 &        & 3 & 1.47 & 41 & .251 & 0.23 \\
                 &        & 4 & 1.03 & 41 & .386 & 0.16 \\
                 &        & 5 & 0.18 & 41 & .859 & 0.03 \\
\hline
Stationary First & Stationary & 1 & 11.61 & 41 & $<$.001 & 1.79 \\
                 &            & 2 & 1.72 & 41 & .154 & 0.27 \\
                 &            & 3 & -0.37 & 41 & .892 & 0.06 \\
                 &            & 4 & 3.52 & 41 & .003 & 0.54 \\
                 &            & 5 & -0.06 & 41 & .955 & 0.01 \\
\hline
\end{tabular}
\caption{One-sample t-tests of win history weights against zero.}
\end{table}

\begin{table}[ht]
\centering
\begin{tabular}{llrrrr}
\hline
Order & Lag & $t$ & $df$ & $p_{\text{FDR}}$ & $d$ \\
\hline
Active First & 1 & -4.49 & 47 & $<$.001 & 0.59 \\
             & 2 & -3.69 & 47 & .001 & 0.45 \\
             & 3 & -0.99 & 47 & .410 & 0.18 \\
             & 4 & -0.48 & 47 & .636 & 0.09 \\
             & 5 &  1.28 & 47 & .347 & 0.23 \\
\hline
Stationary First & 1 & -0.02 & 41 & .988 & 0.00 \\
                 & 2 &  1.99 & 41 & .132 & 0.28 \\
                 & 3 &  2.20 & 41 & .132 & 0.30 \\
                 & 4 & -1.51 & 41 & .230 & 0.30 \\
                 & 5 &  0.20 & 41 & .988 & 0.04 \\
\hline
\end{tabular}
\caption{Paired t-tests of win weights.}
\end{table}

\end{document}